\documentclass[preprint2]{aastex}

\shorttitle{N-body}
\shortauthors{Gauthier \& Martel}

\begin{document}


\title{Long-term stability of collisionless equilibrium configurations;
algorithms}


\author{Jean-Ren\'e Gauthier\altaffilmark{1,2} and 
Hugo Martel\altaffilmark{2,3}}

\altaffiltext{1}{Department of Astronomy \& Astrophysics, 
University of Toronto, Toronto, ON, M5S 1A7, Canada}

\altaffiltext{2}{D\'epartement de physique, g\'enie physique et optique, 
Universit\'e Laval, Qu\'ebec, QC, G1K 7P4, Canada}

\altaffiltext{3}{Department of Astronomy,
University of Texas, Austin, TX, 78712, USA}



\begin{abstract}
We present the summary of the theoretical aspects and algorithms used in an undergraduate (JRG) summer project 
based on numerical N-body simulations of collisionless systems. 
First, we review the importance of numerical N-body simulations in astrophysics. We introduce the different codes used and their performances. 
We then introduce four famous density profiles : Hernquist, NFW, truncated isothermal sphere, and Plummer. The history of these profiles and 
their dynamical properties are discussed in the third section. In the fourth section, we present the Barnes \& Hut tree code, its features and performances.
We then explain how to build and incorporate a multiple time stepping scheme in a tree code. The fifth section is dedicated to the different physical measurements we used to characterize the dynamics of the N-body system. Finally, we describe the future work that will be done with these codes, mainly the study of the adiabatic growth of a black hole at the center of a spherical distribution of stars.    
\end{abstract}




\section{Introduction : numerical N-body simulations in astrophysics}
Numerical simulations, and particularly N-body simulations,  play an important r\^ole 
in astrophysics and cosmology. Before the advances in computer developments, it was impossible
for astrophysicists to compute the evolution of a large system of particles evolving 
under their gravitational influence. The amount of calculation growing like $O(N^2)$, the direct
summation for more than a modest number of particles is unimaginable. 
In 1941, Holmberg \citep{hol41} did a pioneering study of the tidal disturbances due to a close 
encounter between two nebulae. He studied the tidal deformations by reconstructing the orbits of 
each mass elements. He used light bulbs to represent these elements (replacing gravitation by light). The power 
of the bulbs is proportional to the mass and the total light was measured by a photocell. Each nebulae was represented
by 37 light bulbs. Even with this primitive but quite ingenious setup, the creation of spiral arms was detected and a good 
estimate of the energy loss was also provided. 

During the 60's and 70's with the development of computers, scientists
developed direct summation codes to evaluate the potential exerted on a few hundred particles. 
In 1970, Peebles studied the collapse of a cloud of $300$ particles as a model of cluster formation (see \citealt{kly96}). 
Six years later, White studied the collapse of $700$ particles with different masses. Unfortunately, with these low
resolution simulations, numerical artifacts (two-body relaxation) are important and the results from these simulations
were not reliable. The first cosmological simulations were made in the mid 70's. 
Most of these simulations of structure formation confirm the existence of a flat spectrum of initial fluctuations.

The last thirty years have seen the development of several N-body codes. Each one is particularly suitable to study a specific 
range of dynamical problems. Each algorithm has its strengths and weaknesses. In the next few paragraphs, we describe
the characteristics of a few popular codes for collisionless systems, i.e. the study of internal dynamics of galaxy, 
interactions between galaxies and the clustering in an expanding universe \citep{sel87}. 

\subsection{Direct summation codes or particle-particle method}
In this algorithm, the force is computed exactly for each particle. If the system consists of $N$ particles, 
the number of operations to evaluate the force exerted on these particles is proportional to $N^2$. 
This method is flexible, but has a high computational cost and is limited to systems with $N \approx 1000$.  
Usually, the system is integrated using a second-order scheme like leapfrog or Runge-Kutta. The developers 
of these codes were aware of the fact that when particles were closed to each other, the acceleration becomes 
important and one might use multiple timesteps (see section 3) in the integration scheme in order to compute
the particles trajectories accurately. 

\subsection{Grid Methods (Particle-Mesh)}
This method is the fastest when one has to deal with large $N$. The number of computation is of order $O(N + N_g\log N_g)$
where $N_g$ represents the number of grid points. This method determines the force acting on each particle 
by evaluating the gravitational field. The trick is to associate particles to nearby point mesh. Hence, each mesh point has
a defined density attribute. The gravitational potential is evaluated by solving the Poisson equation
\begin{equation}
\nabla^2 \phi = 4\pi G\rho
\end{equation}
on a number of grid points within a fixed volume in space. The force on each particle is obtained via
interpolation techniques. 

There are several ways to assigning particles to the mesh, but one must be careful with the fluctuations of the force when 
the particles are close to each other. The continuity of the derivatives the function used to assign particles to mesh is 
the first criterion. 

The simplest assignment scheme is the ``Nearest-Grid-Point'' (NGP) scheme. The density at a mesh point is determined by
the number of particles in the cell centered on the point divided by the volume of the cell. However, the forces
are discontinuous and this technique is rarely used. There are other more accurate 
assignment schemes. Two of them
are the ``Cloud-in-Cell'' (CIC) and 
``Triangular-Shaped-Cloud'' (TSC) . The former one gives a continuous force, but the first derivative is 
discontinuous. The latter is more accurate. It employs an interpolation function that is piecewise quadratic and each particle is
assigned to a larger number of mesh points. 

The grid methods have been successful in simulations of clustering in an expanding universe and in studying galactic disk dynamics.      
However, these techniques have several limitations. First, the spatial resolution is roughly the distance between mesh points. 
The geometry imposed by the grid cannot be adjusted to the changing shape of the system.  Also, the evaluation of the potential
on grid points where there are no nearby particles is a waste of CPU time and a weakness of the method. 

\subsection{Tree Codes}     
The goal of a tree code is to regroup particles to reduce the number of contribution to the force acting on any particle. 
Basically, this technique has a lower computational cost than the direct summation scheme. The calculations grow like $O(N\log N)$. 
In this scheme, the contributions from nearby particles have to be summed directly, but the particles that are far away can 
be regrouped and replaced by a monopole term in the summation. At each step, the``grouping" can be different because the particles
move in space. Hence, one must rebuild the tree at each time step. This is one drawback of the method. The first two well-known 
tree codes were designed by Appel in 1985 and Barnes \& Hut in 1986. Section 3 is dedicated to the Barnes \& Hut tree code.  

This technique has advantages and weaknesses. First, the tree structure takes into account the shape of the system \citep{bar86}. There is 
no fixed geometry. Second, there is no time lost in evaluating the potential in regions devoid of matter. 
However, this technique is slower than PM methods. Also, it is more difficult to implement.  

Finally, the method has been particularly successful in studying the collisions between galaxies, the internal galactic structures and 
cosmological structures formation. 

\section{Models}
The previous paragraphs introduced basic concepts related to numerical $N$-body techniques. In this section, we present the main physical quantities of four different spherical models used and/or discovered in numerical simulations : Hernquist, NFW, 
Plummer, and TIS density profiles. For each one of them, we include the cumulative mass function, the gravitational potential and the 
distribution function if an approximation or analytical expression can be found.  

\subsection{The Hernquist density profile}
\citet{her90} first introduces this density profile as a fit to the 
de Vaucouleurs surface brightness profile,
\begin{equation}
\frac{I(R)}{I(R_e)} = \exp \bigg[-7.67 \Big( \frac{R}{R_e} \Big) ^{0.25} \bigg] -1
\end{equation}
where $I(R_e)$ and $R_e$ represent respectively half of the total 
luminosity and the isophote value corresponding to it. The de Vaucouleurs profile 
is a good fit for the surface brightness of most elliptical galaxies. It is also a good
approximation for the distribution of stars in the bulge of spiral galaxies. 
The profile has the following shape,
\begin{equation}
\rho(r) = \frac{M}{2\pi}\frac{a}{r}\frac{1}{\left(r+a\right)^3} \; .
\label{hernquist}
\end{equation}
$M$ is the total mass of the system and $a$ is a scale length.
The potential can be found via the Poisson equation. Using the preceding equation, one finds
\begin{equation}
\phi(r) = -\frac{GM}{r+a} \; .
\label{pothernquist}
\end{equation}
It is possible to express the density in function of the potential $\phi(\rho)$
This enables us to evaluate analytically the distribution function (DF) \citep{bin87}
of the system,
\begin{equation}
f(\epsilon) = \frac{1}{ \sqrt{8} \pi^2}\frac{d}{d\epsilon}\int_{0}^{\epsilon}\frac{d\rho}{d\Psi}\frac{d\Psi}{\sqrt{\epsilon - \Psi}} \; .
\label{distributionfunction}
\end{equation} 
The solution for the Hernquist's profile is  
{\setlength\arraycolsep{2pt}
\begin{eqnarray}
f(E) &=& \frac{M}{8\sqrt{2}\pi^3 a^3 v_g^3}\frac{1}{\left(1-q^2\right)^{5/2}}\times {} \nonumber \\
& &{} \times [ 3\sin^{-1}q + q(1-q^2)^{1/2}(1-2q^2) \times {} \nonumber \\
& &{} \times (8q^4-8q^2-3)] \,
\label{hernquistdf}
\end{eqnarray}}
where 
\begin{equation}
q = \biggl(-\frac{aE}{GM}\biggr)^{1/2}
\end{equation}
and
\begin{equation}
v_g = \biggl(\frac{GM}{a}\biggr)^{1/2} \; .
\end{equation}
Useful approximations can be made on this distribution to find the limiting case when $q \rightarrow 0$ 
and $q \rightarrow 1 $. From equation~(\ref{hernquist}) and (\ref{pothernquist}) one can find the the velocity 
dispersion by solving the Jeans equation. For an isotropic system [$\sigma_r = \sigma_{\theta} = \sigma(\phi)$]
where $r \ll a$, 
\begin{equation}
\sigma_{r}^{2} \approx \frac{GM}{a}\frac{r}{a}\ln\frac{a}{r} \; . 
\label{hernveldisp}
\end{equation}
The model provides a good fit for the de Vaucouleurs $R^{1/4}$ 
law [see Fig. 4 in Hernquist (1990)]. Its distribution function [eq.~(\ref{hernquistdf})] is found analytically. 

\subsection{The Plummer profile : a model for the distribution of stars in globular clusters}
The Plummer model \citep{plu11} has been widely used in numerical simulations of star cluster dynamics (e.g. \citealt{aar74}). 
It is a polytrope of index 5 given by : 
\begin{equation}
\rho(r) = \frac{3}{4\pi}\frac{M}{R^3}\frac{1}{\left[1+(r/R)^2 \right]^{5/2}} \; .
\label{plummer}
\end{equation}
where $M$ represents the total mass of the cluster and $R$ is a scale length. The mass enclosed in a radius $r$ is determined by
the following equation : 
\begin{equation}
M(r) = M\frac{r^3}{R^3 \left[ 1+(r/R)^2 \right]^{3/2}}
\label{massplummer}
\end{equation} 
and the gravitational potential is given by
\begin{equation}
U(r) = -\frac{GM}{R}\frac{1}{\left[ 1+ (r/R)^2 \right]^{1/2}} \; . 
\label{potplummer}
\end{equation}
The density profile can be expressed in terms of the gravitational potential. The distribution function 
[see eq.~(\ref{distributionfunction})]
can be found analytically : 
\begin{equation} 
f(E)=\cases{\displaystyle
\frac{24\sqrt{2}\pi^3}{7}\frac{R^2 (-E)^{7/2}}{G^5 M^4}\,, & $E \leq 0$\,; \cr\noalign{\medskip}
       0\,,                                                & $E > 0$\,.\cr}
\label{dfplummer}
\end{equation}

The Hernquist and Plummer models can be easily used as initial conditions for N-body simulations. \citet{aar74}
used a simple method when it is possible to express analytically the radius in function of the mass : 
\begin{equation}
M(r) \to r(M) \; .
\end{equation}  

The technique needs the creation of a subroutine which generates five random numbers $X_1$, $X_2$, $X_3$, $X_4$ and $X_5$
(uniform deviate between 0 and 1). Assume that 
$M(r\to \infty)$ is normalized and $G = M = R = 1$, then $r_j$ (the radius of the $j^{\rm th}$ particle) is determined by $r_j(X_1)$. In the case of 
the Plummer model, 
\begin{equation}
r_j = (X_1^{-2/3} -1)^{-1/2} \; .
\label{radiusdetermination}
\end{equation} 
The $x_j$, $y_j$ and $z_j$ components can be evaluated using $X_2$ and $X_3$ using the following equations : 
\begin{eqnarray}
z_j &=& (1 - 2X_2)r_j \nonumber \\
x_j &=& (r_j^2 - z_j^2)^{1/2}\cos (2\pi X_3) \\
y_j &=& (r_j^2 - z_j^2)^{1/2}\sin (2\pi X_3) \nonumber \; .
\label{xyzdetermination}
\end{eqnarray}

For the velocity, one has to evaluate the escape velocity ($V_e$) at each $r_j$. We can write $V_j/V_e = q_j$ and insert this ratio into 
equation~(\ref{dfplummer}). The probability distribution of $q_j$ is then proportional to 
\begin{equation}
g(q_j) = q_j^2(1-q_j^2)^{7/2}
\label{velocitydf}
\end{equation}
(note : the factor $q_j^2$ comes from the fact that we use the \emph{speed} distribution instead of the velocity component distribution). The values of $q_j$ range between
0 and 1 and $g(q_j)$ is always less than 0.1. By using $X_4$ and $X_5$, we can set $q_j = X_4$ if $X_4$ and $X_5$ fulfill simultaneously the following
criterion : 
\begin{equation}
0.1X_5 < g(X_4) \; . 
\label{velcriterion}
\end{equation}
The velocity components are determined using the same trick as for the position. 
 
\subsection{NFW profile as a result of cosmological simulations}
The Navarro-Frenk-White (hereafter NFW) profile \citep{nav95, nav96, nav97} is 
the profile resulting from the formation of cold dark matter halos in a hierarchical
clustering universe. NFW find that the halo profiles have the same shapes, independent
of the halo mass, initial density fluctuation spectrum and the values of the cosmological 
parameters. 

The profile has the following shape \citep{nav97} : 
\begin{equation}
\frac{\rho(r)}{\rho_{\rm crit}} = \frac{\delta_c}{r/r_s \left( 1+ r/r_s \right)^2}
\label{nfw}
\end{equation} 
where $r_s$ is a scale radius, $\delta_c$ is a dimensionless density and $\rho_{crit} = 3H^2/8\pi G$ corresponding to the critical density for a closed universe. 

The value of $\delta_c$ is strongly correlated with the value of the halo mass. Less massive
halos have higher $\delta_c$, indicating a higher redshift of collapse. 

From the value of the halo ``concentration", $c= r_{200}/r_s$ ($r_{200}$ is the halo radius enclosing a density equals to $200 \rho_{\rm crit}$), it is possible to establish a relation between $c$ and $\delta_c$ : 
\begin{equation} 
\delta_c = \frac{c^3}{\ln (1+c) - c/(1+c)} \; .
\label{deltac}
\end{equation}

The velocity curve is given by :
\begin{equation}
\frac{V_c(r)}{V_{200}} = \frac{1}{x} \frac{\ln(1+cx) - cx/(1+cx)}{\ln(1+c)-c/(1+c)}
\label{nfwvcirc}
\end{equation}
and the cumulative mass function is 
\begin{equation}
M(r) = 4\pi \delta_c \rho_{\rm crit} r_s^3 \bigg[ \ln \Big( \frac{r}{r_s} - 1 \Big) - \frac{r}{r+r_s}  \bigg] \; .
\label{nfwcummulativemass}
\end{equation}

The gravitational potential can be found using equation~(2-22) in Binney \& Tremaine and is equal to
\begin{equation}
\Phi(r) = -\frac{GM(r)}{r} - \frac{4 \pi G \delta_c \rho_{\rm crit} r_s^3}{r+r_s}
\label{nfwpotential}
\end{equation}
where $M(r)$ is given by equation (\ref{nfwcummulativemass}). 

The next step is to evaluate the distribution function of the NFW profile using equation~(\ref{distributionfunction}). The calculation of $d\rho /d\Phi$ shows that $\rho$ cannot be expressed as a function of $\Phi$, the gravitational potential : 
\begin{equation}
\frac{d\rho}{d\Phi} = \frac{\delta_c \rho_{\rm crit} r_s^3 (3r + r_s)}{GM(r)(r+r_s)^3} \; .
\label{drhodphi}
\end{equation}

For NFW, equation (\ref{distributionfunction}) cannot be solved analytically, we need to use numerical methods. However, some authors (e.g. \citealt{wid00}) developed useful approximations for $f(E)$. \citet{wid00} proposed the following fitting formula for the isotropic DF : 
\begin{equation}
F(\epsilon) = F_0 \epsilon^{3/2}(1-\epsilon)^{-\lambda} \left( \frac{-\ln \epsilon}{1-\epsilon}
\right)^q e^P 
\label{widrowdf}
\end{equation}
where $\epsilon = -(E-\Phi_{\infty})/4\pi G \delta_c \rho_{\rm crit} r_s^2$, $\lambda$, $F_0$, and $q$  are the fitting formula parameters for isotropic models. $P$ is a polynomial introduced to improve the fit ($P = \Sigma_i p_i\epsilon^i $) In the case of NFW, $F_0 = 9.1968\times10^{-2}$, $q=-2.7419$, $\lambda = 5/2$, $p_1 = 0.3620$, $p_2 = -0.5639$, $p_3 = -0.0859$, and $p_4 = -0.4912$.

\begin{figure}
\centering{
\includegraphics[scale=0.35]{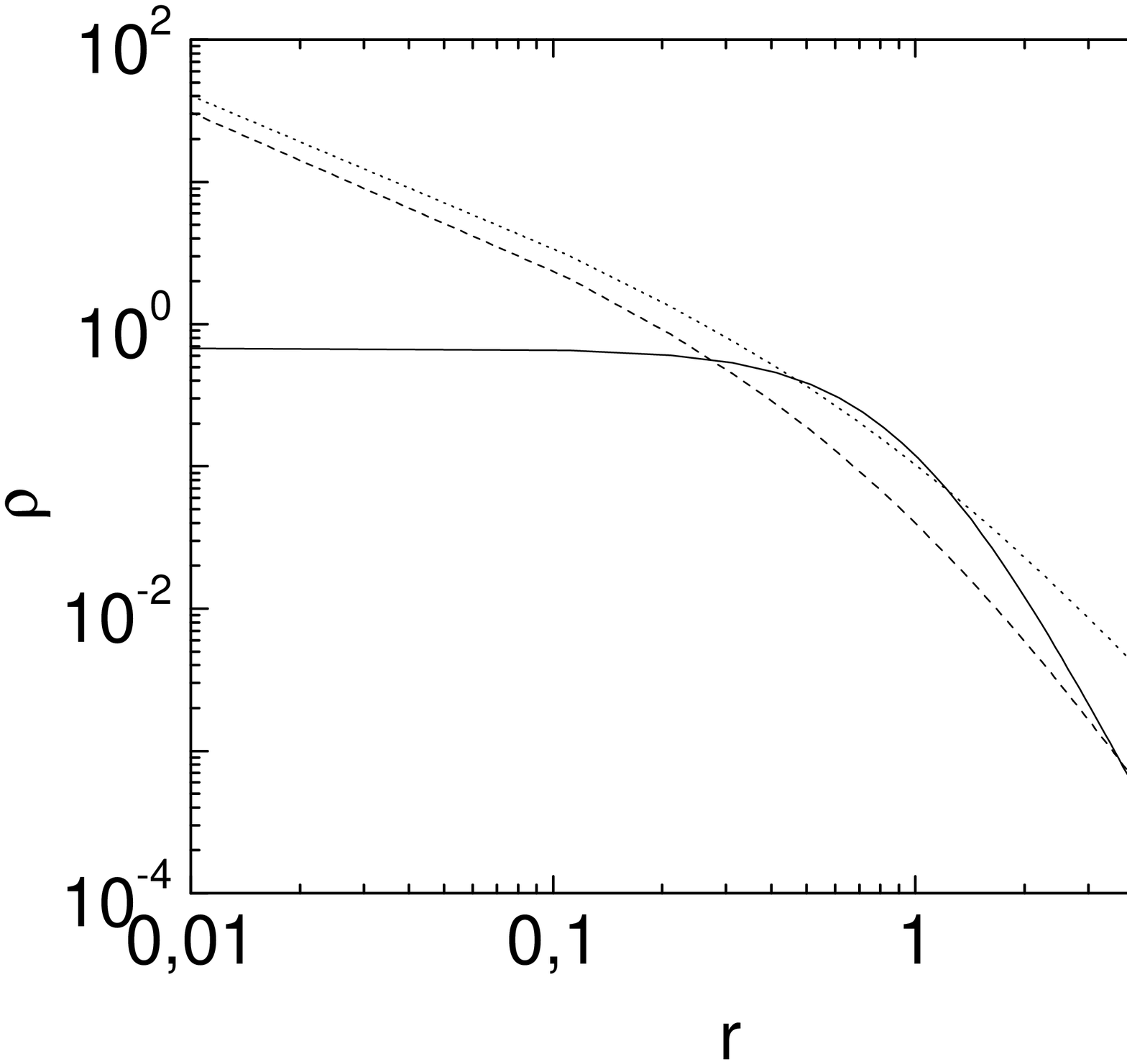}
\includegraphics[scale=0.35]{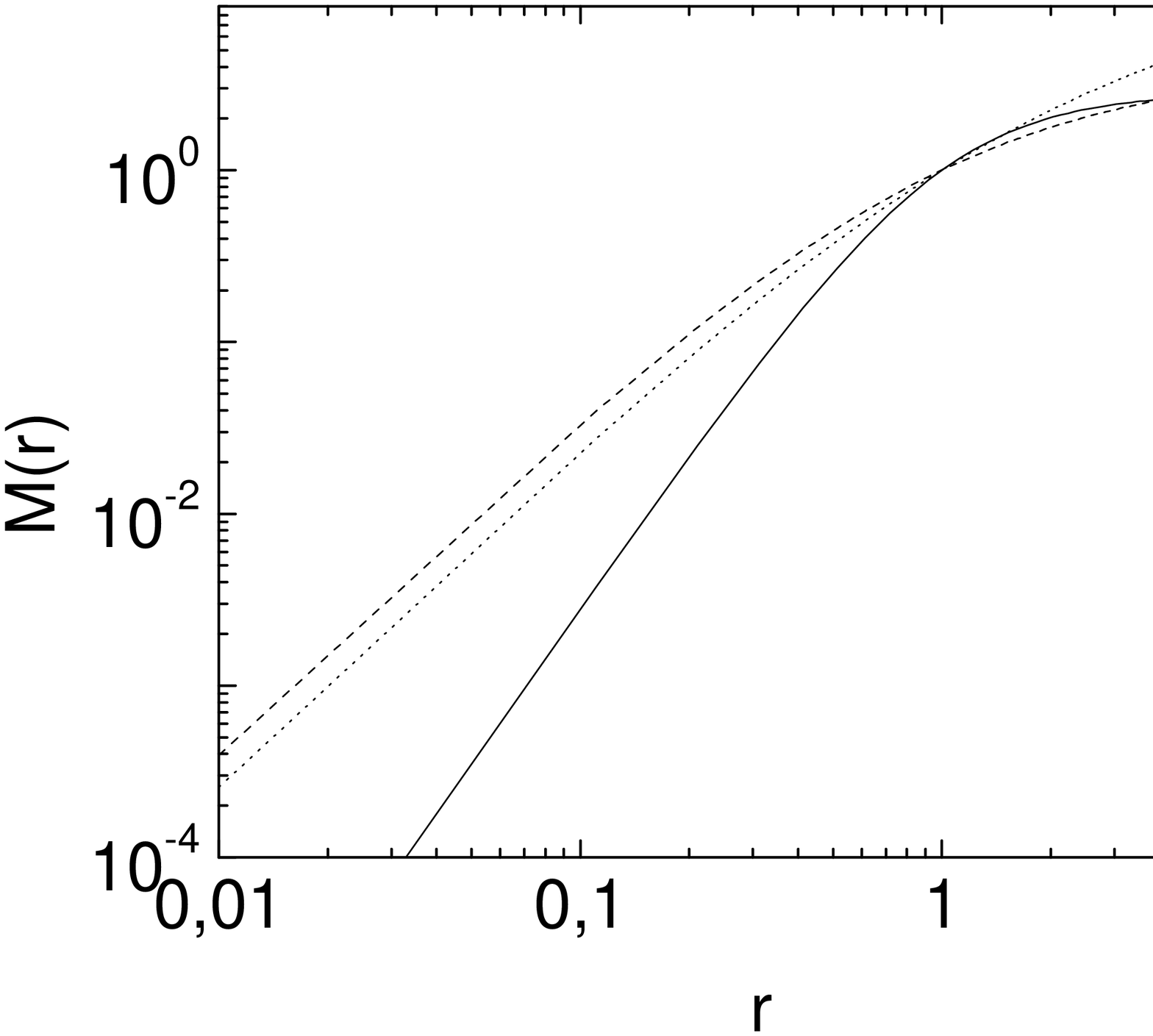}\\
\includegraphics[scale=0.35]{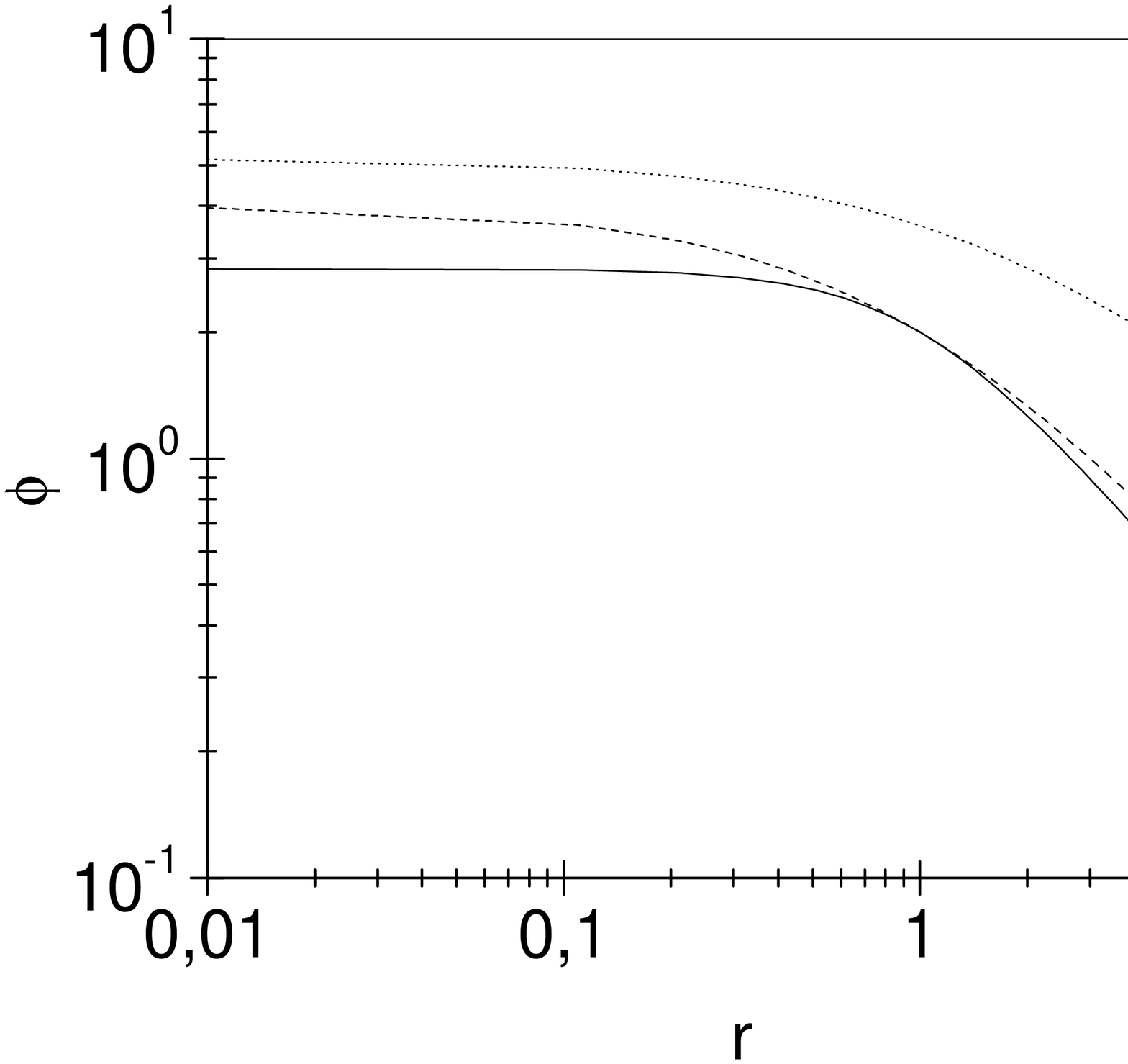}}
\caption{Density, mass and gravitational potential for three models: 
\emph{solid line}: Plummer model; \emph{dashed line}: Hernquist Model; \emph{dotted line}: NFW Model.
For each profile, the scale radius is equal to unity and the normalization is such that $M(r=1) = 1$}
\label{parameters}
\end{figure}

In Figure \ref{parameters}, we give a comparison between the Hernquist, Plummer, and NFW profile for the mass function, 
density profile, and gravitational potential. 

\subsection{The Truncated Isothermal Sphere (TIS) profile}
The TIS (Truncated Isothermal Sphere) is a solution of the Lane-Emden  equation 
\begin{equation}
\frac{d^2\omega}{d\xi^2} + \frac{2}{\xi}\frac{d\omega}{d\xi} = e^{-\omega}
\label{lane-emden}
\end{equation}
where $\omega = \ln(\rho/\rho_0)$, $\xi= r/r_0$ and $r_0$ is the core radius : 
\begin{equation}
r_0^2 = \frac{\sigma^2}{4\pi G \rho_0} \; .
\label{coreradius}
\end{equation}
The TIS model corresponds to the outcome of the collapse and virialization of a top-hat density perturbation. 
This model involves a non-singular, truncated isothermal sphere \citep{sha99}. For a given value of
boundary pressure $p_t$ and mass $M_0$, there is a unique value of $\xi_t$ (where $\xi_t = r_t/r_0$ - $r_t$ is the
truncation radius) which minimizes the total energy. This minimum energy solution is the unique TIS solution 
preferred in nature as the outcome and virialization of a sphere in the presence of a fixed external pressure. 
The virialized object has the same total energy as the top-hat before the collapse. Shapiro \& Iliev found 
that this solution implies $\xi = 29.4$. 

There is a useful fitting formula to equation~(\ref{lane-emden}) that is a good approximation within several core-radii 
\citep{nat97} : 
\begin{equation}
\rho(\xi) = \frac{A}{a^2 + \xi^2} - \frac{B}{b^2 + \xi^2} \; .
\label{approxemden}
\end{equation}
For the particular TIS case, 
\begin{equation}
\left(A, a^2, B, b^2\right) = (21.38, 9.08, 19.81, 14.62) \; . 
\label{solnvariables}
\end{equation}

The TIS profile seems to be a better alternative to the NFW profile in many cases. \citet{sha00} showed that 
the projected mass density of the cluster Cl 0024+1654, determined by strong gravitational lensing, is well fitted by a TIS sphere. 
They also found that a central cuspy NFW fitting profile implies a velocity dispersion which is too large by a factor of $\sim 2$ to be
consistent with the measured velocity dispersion. It also appears that the TIS model can be a good fit to the 
density profile of dark matter-dominated dwarf galaxies \citep{ili01}. Finally, the TIS solution for virialized cosmological halos reproduces fairly well the $\rho_0 - \sigma_v$ 
correlation \citep{sha02}. In fact, current data suggest that the central mass densities $\rho_0$ of cosmological halos in the universe are correlated
with their velocity dispersion $\sigma_v$ over a very wide range of $\sigma_v$ \citep{sha02}. For all these reasons, it is important to study 
this density profile. 

The cumulative mass function for equation~(\ref{approxemden}) \citep{nat97} is 
{\setlength \arraycolsep{2pt}
\begin{eqnarray}
M(\xi) &=& 4\pi \Bigg\{ Aa\bigg[\frac{\xi}{a} - \tan^{-1}\Big(\frac{\xi}{a} \Big)-{} \nonumber \\
& & {}- Bb \bigg[\frac{\xi}{b} - \tan^{-1} \Big(\frac{\xi}{b}\Big) \bigg] \Bigg\} 
\label{massfunctiontis}
\end{eqnarray}}
and the gravitational potential is
{\setlength \arraycolsep{2pt}
\begin{eqnarray}
\Phi(\xi) &=& \frac{2\pi G}{r_0}\Bigg[ \frac{1}{r_0} \bigg( \frac{A-B}{r_0} +{} \nonumber \\
& & {} +\frac{Bb - Aa}{\xi}\tan^{-1}\frac{\xi}{a}\bigg) +{} \\
& & {} + A\ln\Big(a^2 + \frac{\xi^2}{r_0^2}\Big) - B\ln\Big(b^2 + \frac{\xi^2}{r_0^2}\Big) \Bigg] \nonumber
\label{tispotential}
\end{eqnarray}}

Equation~(\ref{approxemden}) can be integrated to yield an analytical fitting formula for the TIS rotation curve \citep{ili01} given by 
{\setlength\arraycolsep{2pt}
\begin{eqnarray}
\frac{v(r)}{\sigma_v} = \Bigg\{ A - B + \frac{1}{\xi} \bigg[ bB\tan^{-1} \left( \frac{\xi}{b} \right)-{} \nonumber \\
{} {}- aA\tan^{-1} \left( \frac{\xi }{a} \right) \bigg] \Bigg\}^{1/2}
\label{vrotation}
\end{eqnarray}} 


In summary, we have briefly introduced few basic models used in current numerical simulations of structure formation, galactic dynamics, 
and star clusters. In the next section, we review the BH tree code, the $N$-body code we used to make these systems evolve. 

\section{Barnes \& Hut Tree code}
The Barnes and Hut (hereafter BH) tree code \citep{bar86} is the one we use for the simulations. 
However, we modified its structure to include a multiple timesteps integration scheme (see section 4). 
In this section we describe how the tree code works : its main features and basic related concepts. 

\subsection{The tree building}

The BH technique is based on a hierarchical division of space into cubic cells. At the beginning of the simulation, 
the particles are distributed, one by one, into the root cell. If two particles are in the same cell, then this cell
is divided into eight ``daughter" cells (an oct tree) of equal volume (see Figs. 1 and 2 in \citealt{bar86}). 
The aim of this process is to make sure that at the end of tree building step, two  particles cannot reside in the same cell. 

For each cell with subcells in it, there is an associated pseudo-particle which contains the total mass in the cell and 
located at the center-of-mass of the particles distribution in the cell. 

This building step must be repeated at each timestep because the particles move and they are no longer in the same cell. 
It takes usually less than 10 \% of the overall CPU time to build the tree \citep{her87}.

\subsection{The force evaluation}

The significant difference with direct summation methods (see section 1.1) resides in the approximation made in the force
calculation. We first consider the interaction between a group of particles (represented by a cell of higher level) and a particle. 
We want to use a criterion based on the size of the cell ($l$) and the distance between the center of mass of this cell and the particle ($d$) to verify if we can consider
this group as a \emph{point-mass particle} or if we need to go down to a lower level of the hierarchy and resolve its constituents. This criterion can be written as 
\begin{equation}
\theta > \frac{l}{d}
\label{openangle}
\end{equation}     
where $\theta$ is a free parameter called the ``opening angle" parameter. Therefore, to compute the force on particle $p$, one needs to walk down the tree, 
starting by the bigger cells and check for each cell if equation~(\ref{openangle}) is verified. If it is the case, then add the contribution of the 
pseudo-particle of this cell to the force exerted on $p$. The result of this ``walking'' process is to reduce the number of terms in the force calculation on one particle
from $N$ to $\log N$.  

The value of $\theta$ is chosen at the beginning of the simulation. \citet{her87} mentioned that for a large $\theta$ ($\theta \geq 1.3 $), the BH code can 
unintentionally include particle self-acceleration by not forcing subdivision of cells containing the particle. It is worth noting that
decreasing $\theta$ from $0.9$ to $0.6$ for $10,000$ particles corresponds to increasing the CPU time by a factor $ \sim 2 - 2.5$. The range of 
typical values for $\theta$ is between $0.7$ and $1.0$. This is a good compromise between accuracy and performance. The typical error for $\theta = 0.9$
relative to a direct sum is  $ \sim 1 \%$ \citep{her87}. The error for a fixed $\theta$ decreases as $N$ increases.  

After establishing the opening angle criterion and determining which cells need to be resolved, one needs to calculate the force exerted by either cells or 
close particles. The particles are represented by point masses. However, the force that two point masses exerts on each other becomes large at 
small distances. Therefore, the velocities will change very rapidly and very short timesteps are needed in order to properly determine 
the motion of the particles (see section 4). We can avoid this problem if the particles are extended \citep{bin87}. To compute the force 
exerted by particle (or pseudo-particle) $j$ on particle $i$, one can use the following formula : 
\begin{equation}
\vec{F}_{ij} = \frac{Gm_im_j \left( \vec{x}_j - \vec{x}_i \right)} {\left( \epsilon^2 + |\vec{x}_i - \vec{x}_j |^2 \right)^{3/2}}
\label{force}
\end{equation}
where $\epsilon$ is the \emph{softening length}. The maximum force occurs at $|\vec{x}_i - \vec{x}_j |^2 = \epsilon^2/2 $. This potential is 
called the \emph{softened point-mass potential} \citep{bin87} also referred as plummer-type softening. 

Finally, for $\theta = 1.0$ and a Plummer model (see section 2.3), the tree walk process and force summation take respectively $80$ \% and $10$ \% of 
the total CPU time \citep{her87}.   

\subsection{The integration scheme}
Once the force has been evaluated for each particle, we need to update the position and velocity of the particles. We describe here two different schemes
of order 2 : leapfrog, and Runge-Kutta. The leapfrog technique is used in the BH tree code. 
\subsubsection{The leapfrog integrator}
Suppose that we know the particles positions and velocities at step $n$ and we want to advance 
the system to step $n+1$. This can be done using a 
time-centered leapfrog : 
\begin{eqnarray}
\vec{v}_{i, n+1/2} & = & \vec{v}_{i, n} + \vec{a}_{i,n}\frac{\Delta t}{2}, \nonumber \\
\vec{r}_{i, n+1} & = & \vec{r}_{i,n} +  \vec{v}_{i,n+1/2}\Delta t, \\
\vec{v}_{i, n+1} & = & \vec{v}_{i, n+1/2} + \vec{a}_{n+1}\frac{\Delta t}{2} \nonumber \; .
\label{leapfrog}
\end{eqnarray}
$\Delta t$ is the timestep. The force is computed at the beginning of the timestep and is used to extrapolate the position and velocity at respectively 
$n+1$ and $n+1/2$. 

\subsubsection{The Runge-Kutta method of order 2}
This method is similar to the Leapfrog scheme. However, it requires the storage of an additional variable, namely $\vec{r}_{i,n+1/2}$. The force is determined
at the center of the timestep. The position and velocity at step $n+1$ are extrapolated from the value of the force at the middle of the timestep \citep{for01} :
\begin{eqnarray}
\vec{v}_{i, n+1/2} & = & \vec{v}_{i, n} +  \vec{a}_{i, n}\frac{\Delta t}{2}, \nonumber \\
\vec{r}_{i, n+1/2} & = & \vec{r}_{i, n} +  \vec{v}_{i, n}\frac{\Delta t}{2}, \\
\vec{v}_{i, n+1} & = & \vec{v}_{i, n} +  \vec{a}_{i, n+1/2}\Delta t, \nonumber \\
\vec{r}_{i, n+1} & = & \vec{r}_{i, n} +  \vec{v}_{i, n+1/2}\Delta t \nonumber \; .
\label{rk2}
\end{eqnarray}
Both methods give similar results.

The BH tree code provides a technique for handling a large number of long-range interactions and concentrating the CPU time allowed on local interactions
where more precision is needed. In this iterative process, the number of mathematical operations grows as $N \log N$.

\section{A multi timesteps scheme}
In the previous section, we described the BH tree code features. Now, we introduce a multi timesteps scheme
which can be inserted in a tree code algorithm. This scheme is especially useful for studies of dynamical properties of
halos containing  black holes, SPH simulations of  star formation, etc. The basic idea is that
far away particles evolve less rapidly than particles closer to the dynamical center of the system. The motion
of distant particles can be integrated with a longer timestep. 

\subsection{A measure of the integration time for each particle}
The first thing we need to do is to evaluate the integration time for \emph{each} particle. One of the most commonly 
used criterion is to associate the velocity and acceleration of the particle to its softening length : 
\begin{equation}
\tau_i = \beta \min \left[ \frac{\epsilon_i}{v_i},\left(\frac{\epsilon_i}{|a_i|}\right)^{1/2}\right]
\label{integrationtime}
\end{equation}
where $\tau_i$, $\epsilon_i$, $v_i$ and $a_i$ are respectively the integration time, softening length, velocity and 
acceleration of particle $i$. $\beta$ is a free parameter varying between $0$ and $1$. This is the most straightforward method 
to estimate the integration time needed for each particle. However,  we are looking for a more 
meaningful criterion based on the orbital period of the particles. 

\begin{figure}
\centering{
\includegraphics[scale=0.8]{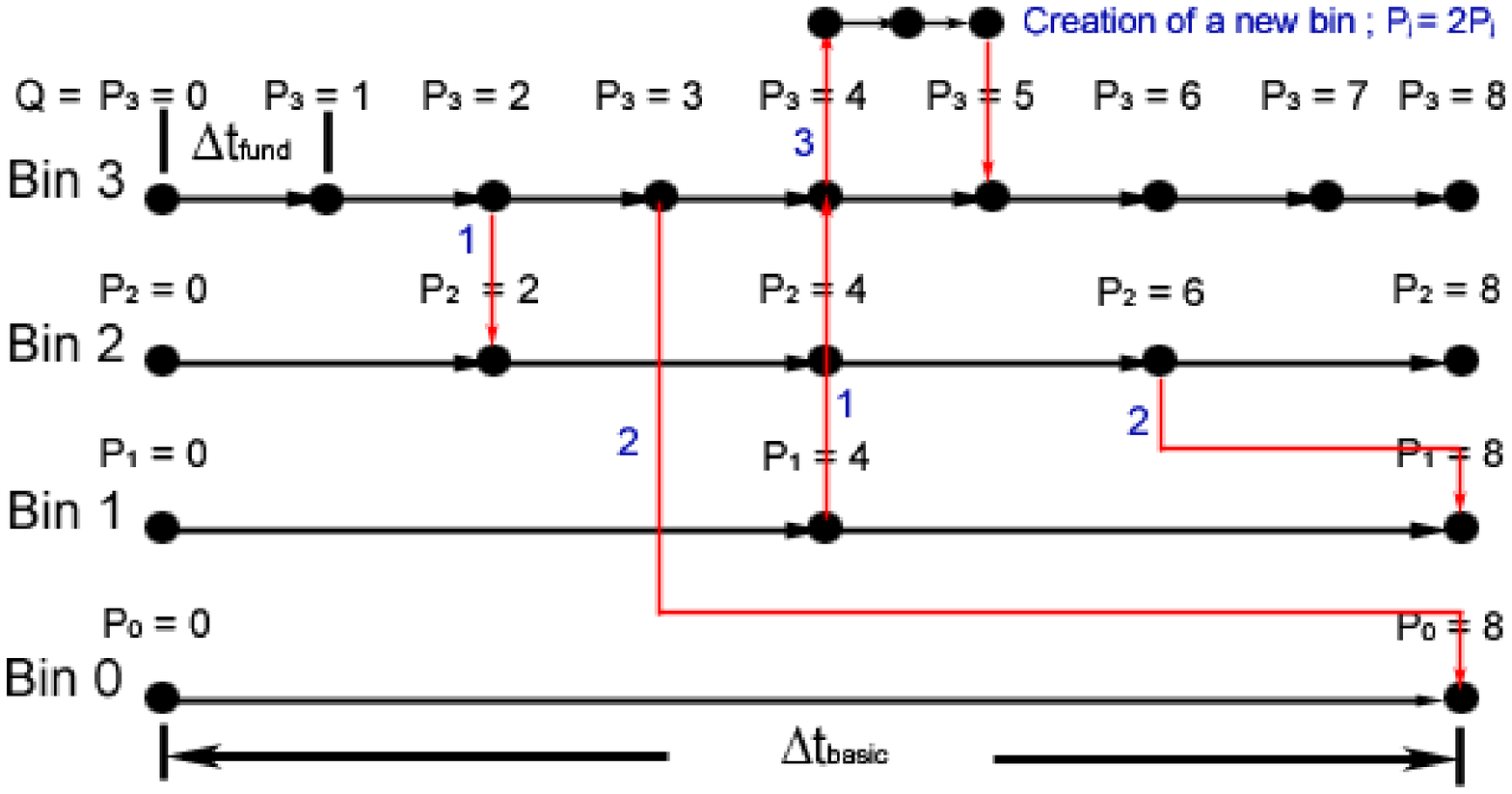}\\
\includegraphics[scale=0.8]{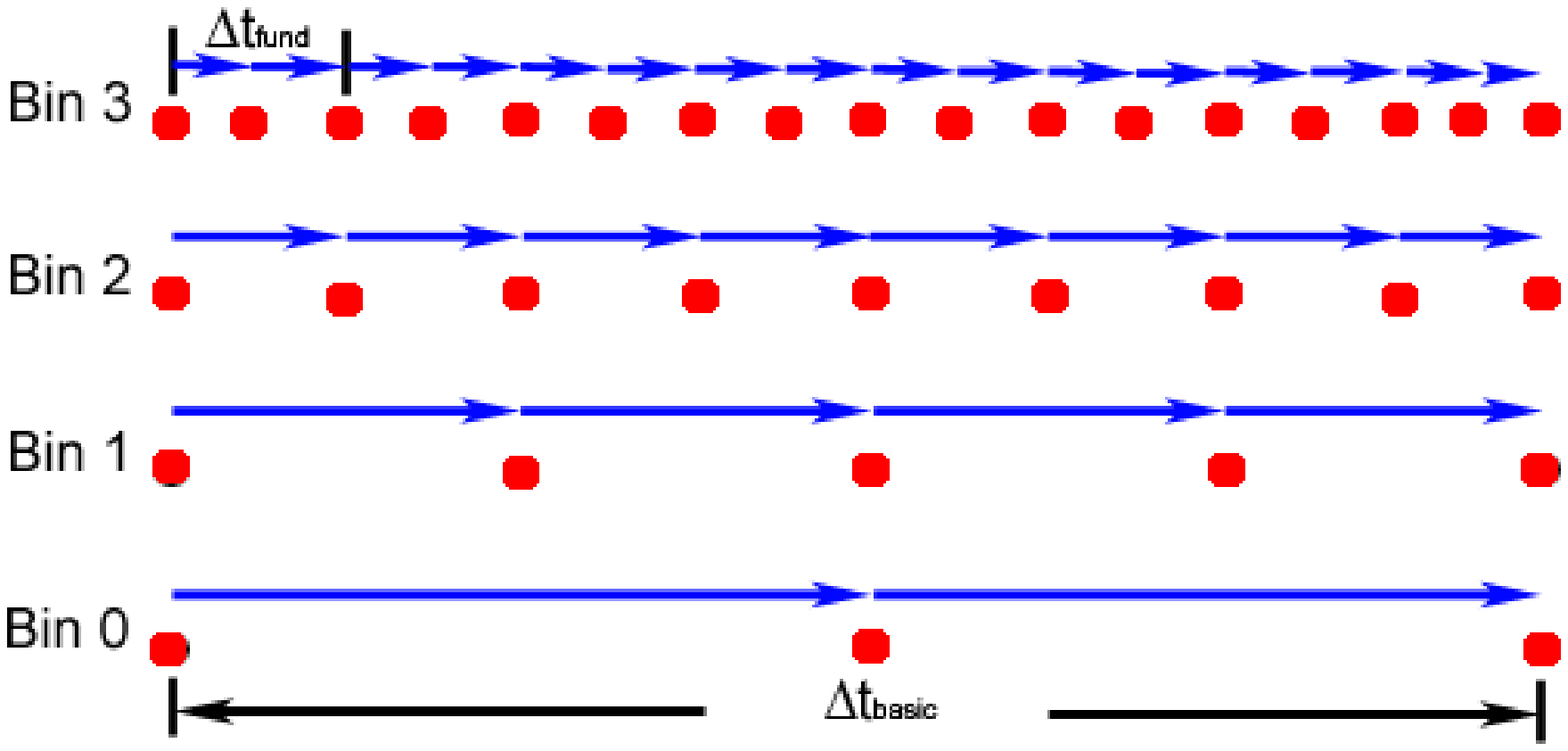}}
\caption{\emph{Upper panel} : A multi timesteps scheme is shown. $P_j$ is computed for each bin. This figure shows
three different bin transitions, as discussed in the text. $1$: These transitions are possible because $P_i = P_j = Q$; 
$2$: These transitions can occur only when $P_i = P_j$. In these cases, the particles have to wait until the bins are
synchronized. $3$: Creation of a new bin. A particle in the fundamental bin needs to decrease its timestep. When the new
fundamental bin is created, $P_j = 2P_j$ for every $j$. \emph{Lower panel}: The red dots indicate where a force 
computation is necessary.}
\label{multi}
\end{figure}

Equation~(\ref{integrationtime}) provides a value of $\tau$ for each particle. The next step is to sort those value of $\tau_i$ ($i  
\in {1,\ldots,N}$) and find $\tau_{\max}$ and $\tau_{\min}$. We then distribute these particles into different bins, having different integration 
timestep. The steps are separated by a factor $2$ in time. If $n$ represents the number of bin, then : 
\begin{equation}
n = 0.69314718\log\left(\frac{\tau_{\max}}{\tau_{\min}}\right)
\label{nbbins}
\end{equation} 
($n$ is round to the next integer). We have now a set of $n$ different timesteps for the $N$ particles. We will use the following notation for
the timesteps : 
\begin{equation}
\Delta t_j, \textrm{\quad where\quad} {j\in 0,\ldots,n-1}
\end{equation}
where $\Delta t_j$ represents the timestep of bin number $j$. The longest timestep, namely $\Delta t_{\rm basic}$ corresponds to $\Delta t_0$ and the shortest
one ($\Delta t_{\rm fund}$) is $\Delta t_{n-1}$. Figure \ref{multi} illustrates the distribution of timesteps for $n=4$. 

\subsection{Evaluating the force acting on each particle}
The goal of introducing a multi timesteps scheme is to reduce the number of force evaluations for the particles having a longer timestep. 
To integrate the motion of the particles, the Barnes \& Hut tree code uses a leapfrog integrator in which the forces are evaluated 
at each half-timestep [see previous section, eq. (35)]. In this multi timesteps scheme, the timesteps are variable between particles and 
therefore, the evaluation of the force on the particles is not necessarily \emph{synchronized} with the evaluation for the other ones. In Figure \ref{multi}, 
we show by red dots, when the force must be evaluated in a system of $n=4$ bins. At the beginning of the simulation ($t=0$), all 
the particles are synchronized. Then, the code updates the position and velocity of each particles (using the leapfrog scheme) with an interval
of $\Delta t_{\rm fund}/2$. At $t=\Delta t_{\rm fund}/2$, only the particles of bin number 3 need a force evaluation and at $t=2\times\Delta t_{\rm fund}/2$,
particles of bin 3 and 2 need a force evaluation. 
There is a simple criterion to check if the bin number $j$ needs a force evaluation. Let $Q$ represents the number of $\Delta t_{\rm fund}/2$ elapses 
since the beginning of the simulations, i.e. $ t = Q\Delta t_{\rm fund}/2$. The $j^{\rm th}$ bin needs a force evaluation if  
\begin{equation}
Q \bmod \frac{\Delta t_j}{2\Delta t_{\rm fund}} = 0  \; .
\label{forcecriterion}
\end{equation} 

It is preferable to use integers in equation~(\ref{forcecriterion}) 
instead of directly comparing $t$ and $\Delta t_j$ because of the possible truncation errors 
modifying the value of $t$ during the simulation. 

\subsection{Particle moving from bin $j$ to bin $i$ }
During the simulation, it is highly probable that many particles will move from one bin to another after evaluating their integration time. 
For example, a particle on a very eccentric orbit will need to diminish its timestep during the closest approach and increase it 
near the pericenter. The multi timesteps scheme
should allow particles to move from one bin to another, but not at any moment. There is a simple criterion to verify if a particle initially
in bin $j$ can move to a bin $i$. First, both bins ($i$ and $j$) must be synchronized, i.e. their internal clock should indicate the same 
time and this time should correspond to the number of fundamental timesteps ($Q/2$) elapses since the beginning of the simulation. 

There is a simple way to evaluate the number of fundamental timesteps elapses for each bin of a set of $n$ bins. 
Suppose that $P_j$ represents the number of fundamental timesteps of the $j^{\rm th}$ bin. $P_j$ will be equal to $Q$ if 
\begin{equation}
Q \bmod \frac{\Delta t_j}{\Delta t_{\rm fund}} = 0 \; .
\label{timetotime}
\end{equation}  
If the following expression is false then $P_j$ keeps its latest value when equation~(\ref{timetotime}) was verified and the particle stays in the same bin.
\begin{equation}
P_i = P_j = Q \; .
\label{movingcriterion}
\end{equation}
You have to wait until equation~(\ref{movingcriterion}) is verified before moving a particle from one bin to another.

\subsection{The value of $n$ changes}
What happens if a particle of the fundamental bin (in our case, bin $n$) goes from bin $n$ to bin $n+1$? Following the previous argument, this 
can happen at each fundamental timestep. In fact, the fundamental timestep changes. It goes from $\Delta t_n$ to $\Delta t_{n+1}$. We must then 
take that into account and change the number of $P$ for each bin. In this example,
\begin{equation}
P_j = 2P_j \textrm{ for }{j \in 0,\ldots,n}
\end{equation} 
and $P_{n+1}$ will be equal to the previous value of $P_n$ multiplied by 2. $Q$ will also be modified, going from $Q$ to $2Q$ and $\Delta t_{\rm fund} = \Delta t_{\rm fund}/2$. 

The opposite phenomenon happens if all the particles of bin $n$ move to bin $n-1$, i.e. we lose a bin. The process is exactly the same except that we divide
by 2 instead of multiplying by 2.

Finally, at the end of a basic timestep, i.e. if 
\begin{equation}
t = m\Delta t_{\rm basic}
\label{reevalutiontime}
\end{equation}  
where $m$ is an integer, we allow all the particles to change from one bin to another after evaluating their integration time.  
Figure \ref{multi} (upper panel) illustrates the different situations examined in the previous paragraphs. 

\section{The measures, observations, dynamical properties of the system}
In the previous three sections, we describe the theoretical aspects of the models we want to study (section 2), the tree code used (section 3) and 
a multi timesteps scheme (section 4) implemented in a leapfrog integrator [see eq.~(35)]. We are interested in determining the 
evolution of the physical properties of the simulated systems.  This section contains a brief description of the different properties of 
the system one can measure using the N-body results. 

\subsection{Conservation of energy}
First, the most important thing to measure when doing N-body experiments is the total energy of the whole system. The energy must be conserved. 
The state-of-the-art simulations are able to conserve total energy within $0.1 - 0.2 \%$. If it is not well conserved this is maybe due to a wrong choice 
of timesteps. The energy is calculated using the following formula : 
\begin{equation}
E = \sum_{i=1}^{N} \sum_{j\neq i}^{N}\Bigg( \frac{1}{2}m_iv_i^2 - \frac{Gm_jm_i}{r_{ij}} \Bigg)
\label{energy}
\end{equation}
where $r_{ij}$ is the distance between particle $i$ and $j$. 

\subsection{Density profile}
The evaluation of the density profile is primordial to study the dynamical properties of a system. In particular, if one studies the collisions between 
galaxies harboring a black hole, the profile can give serious insight about the orbits distribution in the inner few kpc. 

There are basically two ways of measuring the density profile for spherically symmetric systems. The first one is two divide the space into spherical
bins of width $\Delta r$ and count the number of particles in each bin. The bin width can 
increase logarithmically as $r$ becomes large in order to have a good resolution in the 
center and a lower one in the outer parts of the system where nothing important happens. In this picture, the density is evaluated using
the following equation :    
\begin{equation}
\rho_k = \frac{\sum_j N_jm_j}{4\pi r_k^2 \Delta r_k}
\label{density}
\end{equation}
where $\rho_k$, $r_k$ and $\Delta r_k$ are respectively the density, radius and width of bin $k$; $N_j$ represents the number of particles 
of mass $m_j$ in bin $k$. This method has drawbacks when the number of particle is not sufficient to resolve the inner bins. Some bins may 
be empty and others can contain not enough particles. 

The second method consists of sorting the particles by radius,
then divide them by slices of 100 or 1000 particles, for instance, and evaluating 
the average radius, width and density of each slice of particles. This method avoids low resolution bins but one needs to have enough particles
to get slices thin enough for a good resolution.

\subsection{$R_{10}(t)$, $R_{50}(t)$ and $R_{90}$(t)}

In the previous paragraphs, we have seen how to compute the density profile. We are now interested in its evolution. 
$R_{10}(t)$, $R_{50}(t)$ and $R_{90}(t)$ represent the radii which contains
10 \%, 50\% and 90\% of the system mass, respectively. The measurement
of these quantities gives a rough estimate of the evolution of the density profile.

\subsection{Free-fall time}
In the simulations, time is measured in dimensionless units. A good way to characterize the time scales is to express them in terms of the free-fall
time 
\begin{equation}
t_{\rm ff} = \sqrt{\frac{3\pi}{32G\rho}}\,.
\label{freefalltime}
\end{equation}
This quantity represents the time needed for a homogeneous sphere of pressureless material of density $\rho$ (where $\rho$ can be evaluated at $R$ = $R_{50}$) released from rest at $t=0$ to collapse 
to a point at $t=t_{\rm ff}$ \citep{bin87}.   

\subsection{Quadrupole tensor evaluation}
To estimate the size of the cluster, one has to evaluate the components of the quadrupole tensor defined as
\begin{equation}
Q_{ij} = \sum_k m_k \left(r_k \right)_i \left( r_k \right)_j
\label{quadrupole}
\end{equation}
where $m_k$ and $r_k$ are the mass and position relative to particle $k$. The indicies $i$ and $j$ account for the vectors components. 
Once the tensor components have been determined, one has to compute the eigenvalues $Q_1$, $Q_2$ and $Q_3$. The ``semi-major axes" of the 
system are evaluated using the following formula : 
\begin{equation}
a_i = \left( \frac{5Q_i}{M_{\rm tot}} \right)^{1/2} \textrm{ for $i$ } \in {1,2,3} \; . 
\label{semimajoraxes}
\end{equation}     
The ratios of the different $a_i$ give a measure of the system sphericity. If $a_1/a_2 = a_2/a_3 = 1$, then the system is perfectly 
spherical.  

\subsection{Velocity dispersion, anisotropy parameter}
The velocity dispersion for a set of particles is defined as follows: 
\begin{equation}
\sigma_i = \sqrt{\sum_k \frac{[(v_k)_i - \bar{v_i}]^2}{N}}
\label{veldispersion}
\end{equation}
where $\bar{v_i}$ is the average $i$-component of the velocity over the distribution and $N$ is the number of particles. For a spherical 
system, it is useful to find $\sigma_r$, $\sigma_{\theta}$ and $\sigma_{\phi}$. To do this, one needs to find $v_r$, $v_{\theta}$, and $v_{\phi}$ 
These three quantities can be related to $r$, $x$, $y$, $z$, $v_x$, $v_y$, and $v_z$ using the following relations
\begin{eqnarray}
v_r & = & \frac{1}{2} \left(xv_x + yv_y + zv_z\right)\,,  \\
v_{\theta} & = & \frac{r}{\sqrt{r^2 - z^2}}\left[\frac{z}{r^2}\left(xv_x+yv_y +zv_z\right) - v_z\right]\,, \nonumber \\
v_{\phi} & = & \frac{1}{\sqrt{x^2+y^2}}\left(xv_y - yv_x \right)\,. \nonumber
\label{veldispersionspherical}
\end{eqnarray} 
If $\sigma_r = \sigma_{\theta} = \sigma_{\phi}$ then the system is isotropic. 

Another way to measure the degree of isotropy of a system is to evaluate the anisotropy parameter $\beta$ defined as : 
\begin{equation}
\beta = 1 - \frac{\langle v_t^2\rangle}{\langle 2v_r^2\rangle} 
\label{anisotropy}
\end{equation}
where $v_t$ is the tangential velocity. If the distribution of orbits is isotropic, then $\beta = 0$. If the orbits become radial, then 
$\beta \to 1$. For a set of perfectly circular orbits, $\beta \to -\infty$.

\section{Conclusion}
In conclusion, this short review is an tentative introduction to the basic concepts related to 
numerical $N$-body techniques in astrophysics. We introduced different codes used, physical 
models used and developed in numerical simulations, algorithms and measurements. We hope this 
document will be a useful synthesis for any beginner interested in numerical astrophysics. 

\section{Future Work}

This document presents the various basic algorithms needed to do a study of collisionless equilibrium 
systems. Using the techniques presented in this paper, we will study the long-term evolution of density profiles
resulting from the mergers between halos containing black holes.  

\acknowledgments
We acknowledge 
the Natural Science and Engineering Research Council of Canada (NSERC)
for an undergraduate summer fellowship (JRG) and thank
Laurent Drissen and John Dubinski for useful comments during
the project. We also thank Joshua E. Barnes for making his Tree code Guide and
N-body codes available on his website. HM thanks the Canada Research Chair
program for support.

\appendix

\section{Appendix material}
This appendix is dedicated to the description of  an algorithm evaluating the number of possible two-body bound systems which 
can form in $N$-body simulations of dense star clusters. The cluster has to be dense enough so that the softening length is 
short to allow the formation of bound subsystems. The following algorithm is a tentative method to identify possible two-body 
systems in $N$-body files. Those systems can be referred to as ``binary stars". 
\subsection{Brief Overview}
The Binary Stars Finder Code is an analysis tool that can identify the presence of binary systems in a dense cluster of particles.
 To work properly, the
code needs to receive as inputs the dynamical
data of the system, i.e. the position, velocity and acceleration of each particle. Usually this can be done by
using typical output files from N-body codes.                                                                                                                                                             
To be able to identify correctly the formation of potential binary systems, we need to set up conditions
for the formation of such systems.
                                                                                                                                                             
The first one is related to the distance between particles. Suppose that particles A and B
 are constituents of a binary system. Our first hypothesis is that : A is the closest
neighbor to B and vice-versa.
Hence, if there is a third body in the very close vicinity of A, even closer than B,
 we can not consider the system formed by A and B to be a binary one. The probability
that A will interact strongly with this intruder is higher than with B. The condition of reciprocal closest
neighbor must be fulfilled.
                                                                                                                                                             
Second, a cluster consisting of several particles will remain bound if the total energy has
 a negative value. This is also true for binaries. So we have to sum the energy of the two members. If the results respects this second criterion the likelihood that we have identified a real binary increases.
                                                                                                                                                             
The last condition consists of giving a certain level of ``quality" to every system
who already satisfies the first two criteria. This can be achieved by determining the
number of bodies that can potentially disturb the path of one or even two members of the binary
systems. If these intruders are close enough to the considered system, they can break up the whole
system in the next few iterations. In section A.2.3, we describe a method to reduce the problem to a single moving
particle that follows a precised path around a fixed partner. The characteristics of this orbit
will enable us to set up a ``neighborhood of influence" and after that to count the number of ``bad" neighbors.
                                                                                                                                                             
Now the question is how does it work ? This is the object of the next section. In fact, particular features and algorithms will be
discussed. Also, in the last section of this presentation document,
  We suggest several improvements that could be made in further versions of the code.

\subsection{Principal Features}
\subsubsection{``Divide-and-conquer" distribution method}
                                                                                                                                                             
As we saw in the previous section, the first criterion consists of a reciprocal closest
neighbors
selection. To do this, we can simply compute the distance between particle X and every other
 particle of
the cluster and find the closest neighbor to particle X. We could repeat the process for every member of the cluster. Since 
this calculation grows like $N^{2}$,
 for a typical
cluster of $500,000$ that represents a total number of $2.5\times10^{11}$ mathematical operations.
This method has a very high computational cost and should be avoided if possible.
                                                                                                                                                             
Instead of doing a direct calculation between each pair of particle, we can divide the volume that
contains the whole system into individual cubic cells. After that, the comparisons could be
made under the assumption that members of binary systems will
 be close to each other, i.e. members of the same or adjacent cells.
Hence, if we use this method the number of operations ($\Sigma$) will grow approximately like :
\begin{equation}
\Sigma \propto \frac{N^2}{n}
\end{equation}
where $N$ represents the number of particles in the simulation and $n$ the number of cells.
This method is called ``divide-and-conquer.'' Doing it this way we can save precious CPU time.
                                                                                                                                                             
The side length of each cell should be chosen so that it is greater than a few times the average distance between particles (to have the closest neighbor in
the same cell or in an adjacent one). In order to have a significant gain in CPU time, the size of the cell should
be smaller than the size of the whole cluster.
                                                                                                                                                             
\begin{figure}
\centering{
\includegraphics[scale=.80]{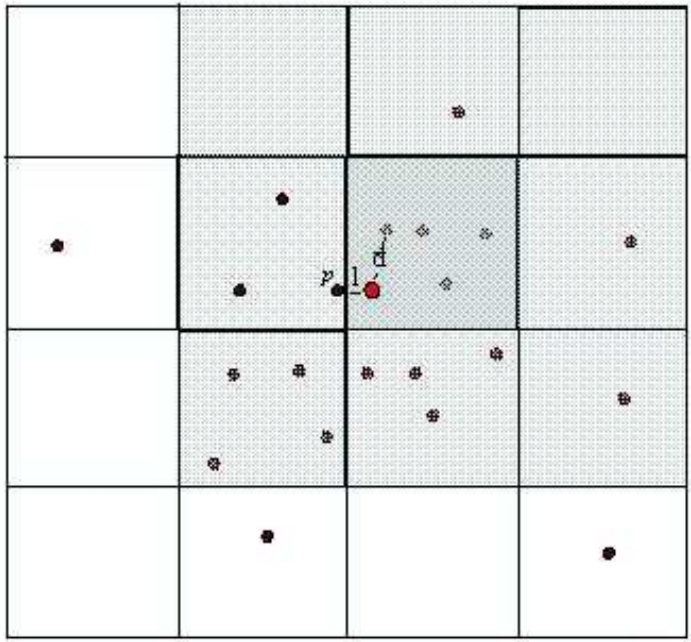}}
\caption{A 2D representation of the division into individual cubic cells. Fade grey cells represent cells adjacent to 
the grey one. $l$ is the distance between the particle for which we want to identifiy its closest neighbor and the wall, 
and $\bar{d}$ is the distance between its closest neighbor in its own cell.}
\label{division}
\end{figure}

Now what happens if the closest neighbor is not in the same cell. This could be possible
 if the particle (particle A) on which we want to identify its closest neighbor is near the
walls of the cell. We can handle this kind of situations by comparing
the distance between A and the closest neighbor of A  in the same cell (value called $\bar{d}$)
with the distance between A and the walls of the cell ($l_{i}$, where i goes from 1 to 6 - a
cube has six faces). Now if
\begin{equation}
\bar{d} > l_{i}
\end{equation}
we must evaluate the distance for particles in the cell labelled $i$.
The process is repeated for adjacent cells who agree to condition (A2). Figure \ref{division} illustrates
the previous considerations.
We can repeat the same process for each one of the 26 adjacent cells of the one we consider.


\subsubsection{Evaluating the Energy of the Bound System}
                                                                                                                                                             
The second step in the identification of binary systems consists of computing the total energy of the two-body system. We can compute 
the total energy of the bound system by using the
following formula:
\begin{equation}
E_{\rm tot} = \frac{1}{2}m_{1}|\vec{v_{1}}|^2 + \frac{1}{2}m_{1}|\vec{v_{2}}|^2 - \frac{Gm_1m_2}{|\vec{r_1}-\vec{r_2}|}
\end{equation}
where the subscripts 1 and 2 denote body number 1 and 2 respectively. Of course, the system will remain bound if $E_{tot}$ has a negative value.
                                                                                                                                                             
\subsubsection{Number of neighbors in the close vicinity of the two-body system}
                                                                                                                                                             
The last step in the identification of binaries consists of evaluating the number of neighbors
in the vicinity of the system. First, we must
characterized the size of the system we are studying. Figure \ref{twobody} is a schematic representation
of a typical two-body system. $\vec{R}$ represents
the center-of-mass position vector. We can reduce this system to a
single particle moving around a fixed massive particle. The motion of this particle must obey to the
specific condition: its angular momentum and energy must be the same as for the previous system and should be conserved along the path.
We can evaluate the reduced angular momentum using this formula :
\begin{equation}
|\vec{L_{\rm tot}}| = |\vec{L_1}|+|\vec{L_2}| = \mu r v = \mu \sqrt{GMa \left(1-e^2\right)}
\end{equation}
where $\vec{r} = \vec{r_2} - \vec{r_1}$ a is the semimajor axis of the elliptic path of the moving particle and $\mu$ is the reduced mass of the two-body system:
\begin{equation}
\mu = \frac{m_1m_2}{m_1+m_2}.
\end{equation}
For a single particle moving around a fixed one, the energy can be written as :
\begin{equation}
E = \frac{1}{2}\mu v^2 - \frac{GM\mu}{r}
\end{equation}
and using the virial theorem:
\begin{equation}
E = \frac{1}{2}\mu v^2 - \frac{GM\mu}{r} = - \frac{GM\mu}{2a}.
\end{equation}
By replacing $a$ in equation (A4) by its value in equation (A7), we can find the ellipse eccentricity $e$ :
\begin{equation}
e = 1 - \frac{2|{\vec L_{\rm tot}}|^2 |E_{\rm tot}|}{\mu^3(GM)^2}.
\end{equation}
Once the eccentricity has been evaluated, we can set the ``influence" radius to be the aphelion distance to the center-of-mass of the system. The aphelion
radius is determined using the following :
\begin{equation}
r_a = a(1+e)
\end{equation}
So, if the distance between a particle $i$ and the center of mass 
of the system is shorter than $\beta r_a$ we can consider this body to be in the close vicinity  (C.V.) of
the system $j$:
\begin{equation}
|\vec{r_i} - \vec{R_j}| < \beta r_{a_j} \to C.V.
\end{equation}
where $\beta$ is a free parameter. It will probably disturb the path of one or maybe two particles.
                                                                                                                                                 


    

\subsection{Further Improvements}
                                                                                                                                                             
Several improvements will be made in a next version of the code. Currently, the code can only handle systems
contained in a cubic volume. Of course, we could put the whole cluster in an augmented cubic volume
but there will be many empty cells and this is not really optimized. Modifying the code so that
systems with rectangular shape can be well-treated is a first thing to do.
                                                                                                                                                             
The current algorithm analyzes only one snapshot data of the system and evaluates the formation of
 binary systems.  It could be really interesting if the code could
integrate the motion of binaries
with the data of several snapshots taken at different times. By doing this, we could be able
to tell if a binary we identified previously is
going to break up or not.

\begin{figure}
\centering{
\includegraphics[scale=.80]{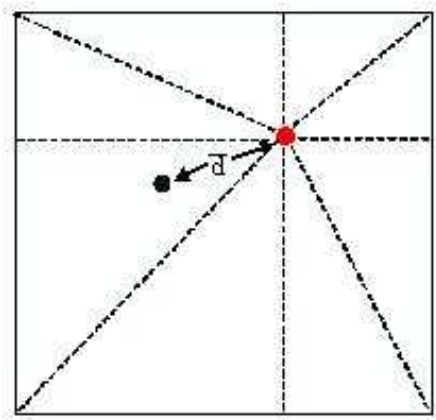}}
\caption{Calculation of the distance between the considered particle, walls, and corners of the cubic cell. In a 2D representation, each square cell has 8 adjacent neighbors.}.
\label{evaluationdistance}
\end{figure}

\begin{figure}
\centering{
\includegraphics[scale=.80]{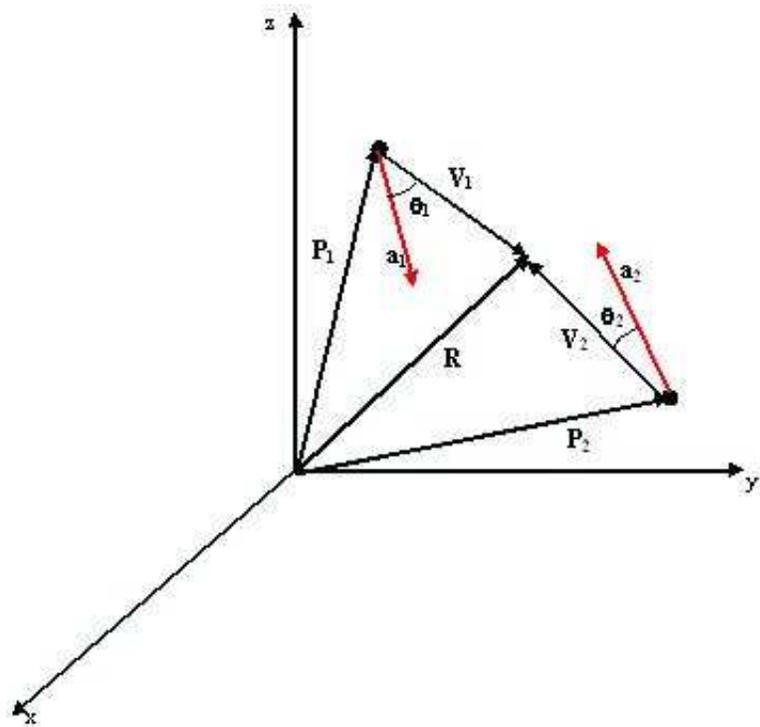}}
\caption{Schematic representation of a two-body system. Vectors $\vec{a}_1$ and $\vec{a}_2$ represent the 
acceleration of particle 1 and 2. respectively.}.
\label{twobody}
\end{figure}

                                                                                                                                                             
\begin{figure}
\centering{
\includegraphics[scale=.80]{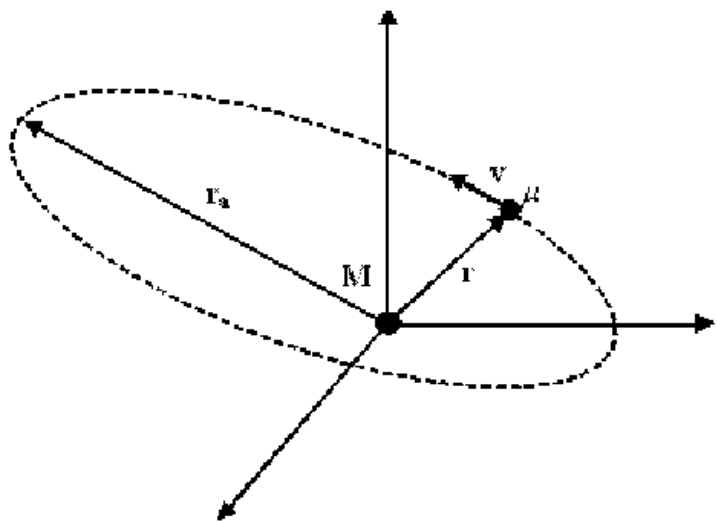}}
\caption{Schema of the reduced system}.
\label{reduced}
\end{figure}

\clearpage

\end{document}